\documentclass[aps,prb,reprint,superscriptaddress]{revtex4-2}

\usepackage{graphicx}      
\usepackage{dcolumn}       
\usepackage{bm}            
\usepackage{hyperref}      
\usepackage{amsmath,amssymb} 
\usepackage{upgreek} 
\usepackage{siunitx} 
\usepackage{comment}

\DeclareSIUnit{\dBm}{dBm}

\begin{document}

\title{Nonlinear frequency shift and bistability of magnon-polarons}

\author{Kevin K{\"u}nstle}
\email{kuenstle@rptu.de}
\affiliation{Fachbereich Physik and Landesforschungszentrum OPTIMAS, Rheinland-Pf{\"a}lzische Technische Universit{\"a}t Kaiserslautern-Landau, 67663 Kaiserslautern, Germany}

\author{Matthias Wagner}
\affiliation{Fachbereich Physik and Landesforschungszentrum OPTIMAS, Rheinland-Pf{\"a}lzische Technische Universit{\"a}t Kaiserslautern-Landau, 67663 Kaiserslautern, Germany}

\author{Philipp Knaus}
\affiliation{Fachbereich Physik and Landesforschungszentrum OPTIMAS, Rheinland-Pf{\"a}lzische Technische Universit{\"a}t Kaiserslautern-Landau, 67663 Kaiserslautern, Germany}

\author{Yannik Kunz}
\affiliation{Fachbereich Physik and Landesforschungszentrum OPTIMAS, Rheinland-Pf{\"a}lzische Technische Universit{\"a}t Kaiserslautern-Landau, 67663 Kaiserslautern, Germany}

\author{Ephraim Spindler}
\affiliation{Fachbereich Physik and Landesforschungszentrum OPTIMAS, Rheinland-Pf{\"a}lzische Technische Universit{\"a}t Kaiserslautern-Landau, 67663 Kaiserslautern, Germany}

\author{Katharina Lasinger}
\affiliation{Fachbereich Physik and Landesforschungszentrum OPTIMAS, Rheinland-Pf{\"a}lzische Technische Universit{\"a}t Kaiserslautern-Landau, 67663 Kaiserslautern, Germany}
\affiliation{Clarendon Laboratory, Department of Physics, University of Oxford, Parks Road, Oxford OX1 3PU, United Kingdom}

\author{Matthias R. Schweizer}
\affiliation{Fachbereich Physik and Landesforschungszentrum OPTIMAS, Rheinland-Pf{\"a}lzische Technische Universit{\"a}t Kaiserslautern-Landau, 67663 Kaiserslautern, Germany}

\author{Philipp Pirro}
\affiliation{Fachbereich Physik and Landesforschungszentrum OPTIMAS, Rheinland-Pf{\"a}lzische Technische Universit{\"a}t Kaiserslautern-Landau, 67663 Kaiserslautern, Germany}

\author{John F. Gregg}
\affiliation{Clarendon Laboratory, Department of Physics, University of Oxford, Parks Road, Oxford OX1 3PU, United Kingdom}

\author{Mathias Weiler}
\affiliation{Fachbereich Physik and Landesforschungszentrum OPTIMAS, Rheinland-Pf{\"a}lzische Technische Universit{\"a}t Kaiserslautern-Landau, 67663 Kaiserslautern, Germany}
\date{\today}

\begin{abstract}
We investigate the nonlinear dynamics of strongly coupled surface acoustic waves (SAWs) and spin waves (SWs) in a magnetoacoustic resonator based on a YIG/ZnO heterostructure by combining microwave reflection measurements with microfocused Brillouin light scattering spectroscopy. In the linear regime, the electrical response reveals clear hybridization between standing SAW cavity modes and finite-wave-vector SWs, resulting in pronounced avoided crossings. At elevated drive powers, the hybrid system exhibits a strongly field-dependent nonlinear response characterized by a positive frequency shift of the driven SW mode. Using the vector Hamiltonian formalism for nonlinear spin-wave dynamics, we show that this shift is dominated by a cross-shift term. In our resonator geometry, this contribution becomes significant because the standing SAW cavity mode simultaneously excites counterpropagating SWs with wave vectors $+k$ and $-k$. For suitable field detuning, the nonlinear shift drives the SW mode into resonance with the SAW excitation, leading to a strong enhancement of the magnon population, broadband nonlinear scattering, and bistable foldover behavior. Beyond the foldover threshold, both the magnon and phonon responses stabilize. These results establish SAW-driven $k \neq 0$ magnon-phonon hybrids as a promising platform for nonlinear magnetoacoustics and wave-based information processing.
\end{abstract}

\maketitle

\section{Introduction}

Hybrid excitations that combine elastic, magnetic, and electromagnetic degrees of freedom provide a route toward compact, energy-efficient platforms for wave-based information processing \cite{Li2020HybridMagnonics,Awschalom2021QuantumEngineering}. In particular, elastic waves can generate and manipulate spin excitations on chip through magnetoelastic coupling \cite{Puebla2022-xx, Bozhko2020-mg}. Surface acoustic waves (SAWs) are especially attractive in this context because they provide an efficient interface between microwave electronics and spin-wave (SW) excitations, enabling coherent and wave-vector-selective driving of magnons in magnetic materials \cite{PhysRevB.86.134415, Kus2020-ks, Kunz2025-cm, Lyons2023-op, Kunz2024-qd}. Over the past years, this platform has been used to demonstrate a wide range of magnetoacoustic phenomena \cite{Li2021-sz,Yang2021-ue}, including the strong coupling of SAWs and SWs~\cite{Hwang2024-bt, natcom2025}, nonreciprocal acoustic transmission \cite{Kus2024-ca}, and tunable magnonic crystals \cite{Kryshtal2012-nm}. As a result, SAW-based magnonics has developed into a versatile framework for studying coherent spin-lattice interactions and for implementing functional concepts in spintronic and magnonic architectures. To date, however, most of these magnetoacoustic phenomena have been investigated in the linear regime, where SAW and SW amplitudes are small and can be treated as perturbations of the equilibrium elastic and magnetic ground states \cite{PhysRevB.86.134415}. 

Beyond the linear regime, magnetoelastic systems are expected to exhibit rich dynamics, mediated in part by the intrinsic nonlinearity of SW systems~\cite{Bruhlmann2026-pm}. These nonlinear magnon interactions give rise to a broad range of phenomena, including nonlinear frequency shifts \cite{Lake2022NonlinearSpectralShiftYIG, Breitbach2024-xu}, foldover instabilities \cite{Wang2024-su, Bunkov2021-np, PhysRevLett.130.046703}, and parametric scattering processes \cite{Heinz2022-qy, Kuhn2026-jp, Jander2025-qd}. Nonlinear magnon interactions are of fundamental interest and may also be exploited for signal processing, threshold devices, and nonlinear information transduction. In this context, SWs constitute a particularly attractive platform, as their nonlinear response is both pronounced and highly tunable by the externally applied magnetic field, geometry, and mode structure \cite{Zheng2023-rr, krivosik_hamiltonian_2010}. Such effects have been studied extensively in directly microwave-driven spin-wave systems \cite{Flebus2024-ka} and in systems based on quasi-uniform magnetic modes coupled to microwave cavity photons \cite{PhysRevLett.130.046703, PhysRevLett.120.057202}. More recently, nonlinear magnetoelastic phenomena driven by SAWs have also begun to attract attention. Hwang \textit{et al.} reported harmonic and subharmonic magnon generation in a SAW resonator~\cite{Hwang2026-ed}, Geilen \textit{et al.} demonstrated parametric excitation and instabilities of spin waves driven by SAWs~\cite{Geilen2025-ai}, and Shah \textit{et al.} investigated the symmetry and nonlinearity of spin-wave resonance excited by focused SAWs~\cite{Shah2023-if}. In addition, Alekseev \textit{et al.} studied linear and parametric spin-wave excitation in a YIG-based bulk-acoustic-wave resonator~\cite{Alekseev2025-ga}. Here, we show that strongly coupled finite-wave-vector SAW-SW hybrid modes in low-damping YIG exhibit foldover-type bistability, nonlinear spectral redistribution, and stabilization of both magnon and phonon responses beyond the foldover.

We investigate this regime using a one-port SAW resonator based on a YIG/zinc oxide (ZnO) heterostructure \cite{Lewis1972-bb, Ryburn2025-fq, manenti_surface_2016}. The measured electrical signal at the resonator port is governed primarily by the SAW response and therefore provides indirect access to the SW dynamics through magnetoelastic coupling. To investigate the involved scattering processes, we combine electrical spectroscopy with polarization-resolved microfocused Brillouin light scattering ($\upmu$BLS), which provides frequency-resolved access to populated magnon and phonon spectra \cite{Sebastian2015-hs, Geilen2022-zo, Buttner2000-ok, Kunz2024-qd}. We observe pronounced nonlinear phenomena, including foldover-type bistabilities and a power-dependent redistribution of the magnon population over a broad frequency range. Notably, beyond the bistability threshold, both the magnon and phonon responses stabilize. By comparing these experiments with calculations based on the vector Hamiltonian formalism for nonlinear spin-wave dynamics \cite{tyberkevych2020vector}, we find qualitative agreement with a model that includes the nonlinearities on the magnonic side, without explicitly treating the full nonlinear dynamics of the coupled SAW resonator. This supports the interpretation that nonlinear magnon interactions play a central role in the observed response. Our results thereby connect the macroscopic nonlinear response of a SAW resonator to the microscopic spectral redistribution of magnons and establish low-damping magnetoelastic YIG devices as a promising platform for nonlinear magnetoacoustics.

The paper is organized as follows. First, we describe the experimental platform and measurement techniques. We then present the electrical response of the hybrid system and introduce the theoretical framework used to analyze the relevant nonlinear spin-wave interactions. Finally, we employ $\upmu$BLS to directly resolve the spectral redistribution induced by the nonlinear interactions and compare the optical and electrical responses in the regime of foldover bistability.

\section{Experimental Methods}

\begin{figure}[tbp]
    \centering
    \includegraphics[width=\columnwidth]{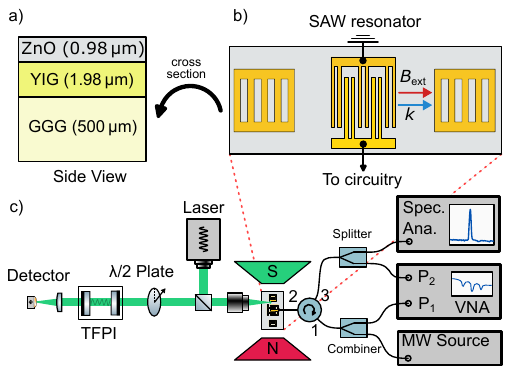}
    \caption{Schematic of the sample and measurement setup. (a) Heterostructure consisting of piezoelectric ZnO deposited on a YIG/GGG stack. (b) SAW resonator patterned on the ZnO layer; the device is identical to that studied in Ref. \cite{natcom2025}. (c) Combined optical and electrical measurement setup. The sample is mounted in BV geometry inside an electromagnet, allowing simultaneous measurements of the electrical reflection and the polarization-resolved $\upmu$BLS response of excited quasi-particles}
    \label{fig:Aufbau}
\end{figure}

Our magnetoacoustic system is based on a high-quality YIG film with a thickness of $\qty{1.98}{\micro\meter}$, grown by liquid phase epitaxy on a [111]-oriented GGG substrate (Fig.~\ref{fig:Aufbau}~(a). To enable electromechanical transduction of SAWs, a piezoelectric ZnO layer (\qty{980}{\nano\meter}) is deposited on the YIG surface by RF magnetron sputtering. A one-port SAW resonator, schematically shown in Fig.~\ref{fig:Aufbau}~(b), was patterned using electron-beam lithography. The device is identical to that characterized in our previous work \cite{natcom2025}.

The resonator comprises a split-finger interdigital transducer (IDT) for SAW excitation and detection, which is flanked by two Bragg mirrors formed by shorted metallic electrode arrays acting as acoustic reflectors \cite{MORGAN2007225}. The resulting Fabry–Pérot-type acoustic cavity supports several high-quality-factor (high-$Q$) standing SAW modes within the first stopband of the acoustic mirrors, corresponding to a wavelength of $\lambda = \qty{2}{\micro\meter}$.

In comparison to a conventional SAW delay line driven with identical microwave power, the resonator geometry enhances the acoustic strain amplitude in the structure \cite{HwangBragg}, which penetrates into the YIG underlayer. This strain generates, via magnetoelastic coupling, a dynamic magnetic driving field \cite{PhysRevB.86.134415} that can induce nonlinear SW dynamics at sufficiently high SAW driving power. Additionally, the standing SAW in the resonator is composed by counterpropagating SAWs, in contrast to the single SAW propagation direction in an acoustic delay line.

To characterize the SAW and SW dynamics, the sample is mounted in an electromagnet that provides an external magnetic field $B_\mathrm{ext}$ oriented parallel to the in-plane SAW wave vector $k$. This configuration corresponds to the backward-volume (BV) SW geometry. Under these conditions, $\upmu$BLS spectroscopy \cite{Sebastian2015-hs} is combined with simultaneous electrical measurements to characterize the excited quasi-particles. A schematic of the combined setup is shown in Fig.~\ref{fig:Aufbau}~(c).

Throughout this manuscript, three characteristic magnetic fields appear repeatedly and are introduced here for clarity. $B_\mathrm{AC}=\qty{34.57}{\milli\tesla}$ denotes the magnetic field corresponding to the center of the anticrossing in the linear power regime. $B_\mathrm{S1}=\qty{32.70}{\milli\tesla}$ and $B_\mathrm{S2}=\qty{32.52}{\milli\tesla}$ denote two smaller magnetic field magnitudes below $B_\mathrm{AC}$, such that $B_\mathrm{S2}<B_\mathrm{S1}<B_\mathrm{AC}$.

The electrical measurement circuit consists of a vector network analyzer (VNA) and an external microwave (MW) source connected to the device via a RF-combiner and a RF-circulator. The signal reflected from the IDT is split and routed simultaneously to port 2 of the VNA and to a spectrum analyzer (SA). Although the VNA records $S_\mathrm{21}$, the wiring of the measurement circuit makes its magnitude equivalent to the conventional $|S_{11}|$ reflection response of a one-port resonator. It is therefore labeled as $|S_{11}|$ in the following. The MW source and SA are used to perform controlled bidirectional power sweeps.

Optical detection of the SAW and SW dynamics is performed using $\upmu$BLS spectroscopy \cite{Kunz2024-qd,Geilen2022-zo}. A continuous-wave laser with a wavelength of \qty{532}{\nano\meter} is focused onto the sample through an objective with a numerical aperture of $\mathrm{NA}=0.75$, resulting in a spot diameter of approximately \qty{430}{\nano\meter} and a maximal detectable wave vector of $k_\mathrm{max} \approx \qty{17.7}{\radian\per\micro\meter}$. A half-wave plate is placed in the beam path in front of the polarization-sensitive tandem Fabry–Pérot interferometer (TFPI). The combination of the half-wave plate and the interferometer enables polarization-resolved detection of the $\upmu$BLS signal, allowing magnon- and phonon-induced contributions to be distinguished. This separation is possible because the scattered photons undergo different polarization rotations in the two scattering processes \cite{Sebastian2015-hs, Geilen2022-zo}.

\section{Results and discussions}

\subsection{VNA reflection measurements}
\begin{figure*}[tbp]
    \centering
    \includegraphics[width=\textwidth]{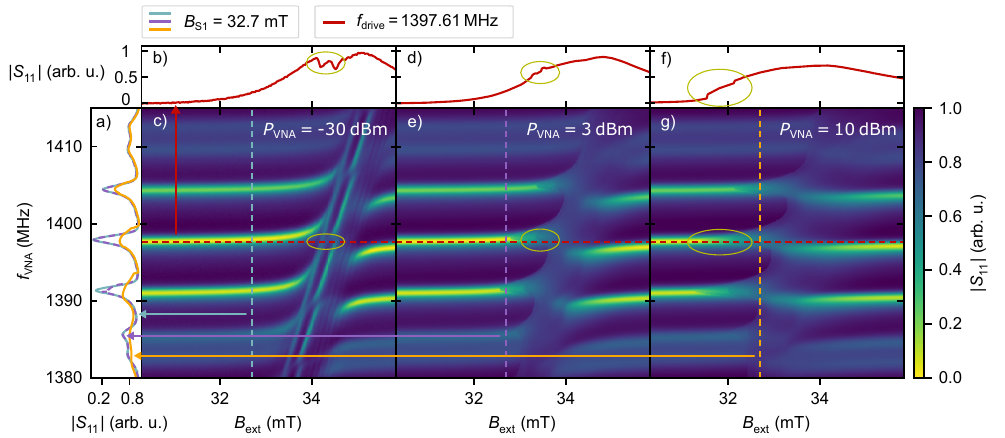}
    \caption{Power dependent VNA reflection $|S_\mathrm{11}|$ measurements. (c,e,g) $|S_\mathrm{11}|$ parameter colorcoded in dependence of $B_\mathrm{ext}$ and $f_\mathrm{VNA}$ for an applied VNA power of $P_\mathrm{VNA}=\qtylist{-30;3;10}{\dBm}$ respectively. Three high-$Q$ SAW modes (horizontal light green features with reduced reflection) are visible and show a pronounced anticrossing with the BV-SW mode at $P_\mathrm{VNA}=\qty{-30}{\dBm}$. With increasing $P_\mathrm{VNA}$, the hybridization branches broaden and move to lower $B_\mathrm{ext}$. In addition, the color-coded $|S_\mathrm{11}|$ data develop step-like features that depend on both magnetic field and frequency.  
    (a) Frequency line cuts at fixed $B_\mathrm{S1}=\qty{32.7}{\milli\tesla}$ (line colors correspond to the dashed lines in panels (c,e,g)). Compared to the low-power case, the resonances broaden and decrease in amplitude at higher powers. Jumps in $|S_\mathrm{11}|$ are visible in the high-power case for the the upper peak flanks of the two high-$Q$ modes lower in frequency.
    (b,d,f) Field line cuts at fixed drive frequency $f_\mathrm{drive}=\qty{1397.61}{\mega\hertz}$ (red dashed line) for the three depicted VNA powers. The olive ellipses mark the regions where additional features occur. At $\qty{-30}{\dBm}$, the drops in the $|S_\mathrm{11}|$ parameter correspond to the upper hybridization branch of the lower high-$Q$ mode and the directly excited BV-SW due to spurious RF currents in the microstructure. At the higher powers, two jumps are visible, at which $|S_\mathrm{11}|$ increases.}
    \label{fig:VNA}
\end{figure*}

We first discuss our VNA reflection measurements, that characterize key experimental observations of the coupled system. As the underlying physical mechanisms cannot be fully inferred from the electrical response alone, we discuss supporting frequency-resolved spectroscopy of magnons and phonons recorded using complementary techniques in the following sections.

The SAW resonator is mounted in an electromagnet to apply an external magnetic field in BV geometry and electrically connected to a VNA.

The wavelength of the driven SAW cavity mode is set by the resonator and IDT geometry, yielding \mbox{$\lambda_\mathrm{drive}=\qty{2}{\micro\meter}$} and \mbox{$k_\mathrm{drive}=\qty{3.142}{\radian\per\micro\meter}$}. Since the SWs are resonantly driven by this SAW mode, their wavelength is likewise fixed by the resonator geometry and is independent of the applied magnetic field.

Figure~\ref{fig:VNA}~(c) shows the linear response of the resonator measured at a VNA power of $P_\mathrm{VNA} = \qty{-30}{dBm}$. Three high-$Q$ SAW modes are visible as horizontal features (light green). Sweeping the external magnetic field $B_\mathrm{ext}$ reveals hybridization of these modes with the BV spin-wave mode (BV-SW), giving rise to pronounced avoided crossings. Using the analysis established in Ref.~\cite{natcom2025}, we extract a coupling strength of $g/2\pi = \qty[separate-uncertainty = true]{5.05(0.16)}{\mega\hertz}$, together with SAW and SW decay rates of $\kappa_\mathrm{p}/2\pi = \qty[separate-uncertainty = true]{1.81(0.02)}{\mega\hertz}$ and $\kappa_\mathrm{m}/2\pi = \qty[separate-uncertainty = true]{1.54(0.03)}{\mega\hertz}$, respectively. Since $g > \mathrm{max}(\kappa_\mathrm{p}, \kappa_\mathrm{m})$, the system is in the strong-coupling regime.

With increasing microwave power [Fig.~\ref{fig:VNA}~(e,g)], the hybridized branches broaden and progressively lose contrast, reminiscent of the behavior reported for the magnon-photon hybridization in microwave cavity-based nonlinear magnon-polariton systems \cite{PhysRevLett.130.046703}. In addition, the SW resonance and hybridization point shifts to lower magnetic fields with increasing drive power. These observations already indicate a pronounced nonlinear modification of the coupled SAW-SW response under strong driving. Furthermore, additional discontinuous features emerge that depend on both $B_\mathrm{ext}$ and $f_\mathrm{VNA}$.

To illustrate the power dependence in more detail, Fig.~\ref{fig:VNA}~(a) shows frequency line cuts taken at a fixed magnetic field of $B_\mathrm{S1} = \qty{32.7}{\milli\tesla}$. The line colors correspond to the cuts indicated in Fig.~\ref{fig:VNA}~(c,e,g). In the linear regime, the three unperturbed high-$Q$ SAW modes are clearly resolved. At elevated powers, the resonances broaden and their amplitudes are reduced. In addition, discontinuous changes in $|S_\mathrm{11}|$ become visible, most clearly on the upper-frequency flanks of the two lower-frequency high-$Q$ modes in the highest-power trace.

The field shift of the coupling features towards lower $B_\mathrm{ext}$ is also apparent in the two-dimensional maps of Fig.~\ref{fig:VNA}~(c,e,g) and becomes more pronounced with increasing power. We attribute this shift to a positive nonlinear frequency shift of the SW mode as its population increases under stronger magnetoelastic driving. Since the resonator geometry fixes the driven SAW wave vector, and thus the wave vector of the coupled SWs, the nonlinear increase in the SW frequency requires a lower $B_\mathrm{ext}$ to maintain resonance. Consequently, the corresponding coupling features shift to smaller magnetic fields. This observation is central to the nonlinear behavior of the system and is analyzed in detail in Sec.~\ref{sec:Nonlinear}.

At the same time, the avoided-crossing features become increasingly difficult to resolve at high power. This indicates that strong driving modifies the effective dynamics of the coupled system, for example through changes in the effective coupling strength and the relevant decay rates. We return to this point in Sec.~\ref{sec:BLS} and \ref{sec:SA} .

Additional insight is provided by the field line cuts shown in Fig.~\ref{fig:VNA}~(b,d,f), taken at a fixed drive frequency of $f_\mathrm{drive} = \qty{1397.6}{\mega\hertz}$, corresponding to the center frequency of the uncoupled high-$Q$ SAW mode. The olive ellipses mark the main regions of interest. In the low-power case [Fig.~\ref{fig:VNA}~(b)], the line cut reproduces the behavior reported in Ref.~\cite{natcom2025}: the left feature within the ellipse corresponds to the hybridized branch of the lower frequency high-$Q$ SAW mode, while the right dip originates from the directly excited SW resonance caused by spurious RF currents in the microstructure.
Apart from these features, $|S_\mathrm{11}|$ increases as $B_\mathrm{ext}$ approaches the ideal resonance condition, displaying the shift of the hybridization relative to the drive frequency.

At higher powers, the onset of the increased $|S_\mathrm{11}|$ shifts to lower magnetic fields, due to the positive nonlinear shift discussed above. Further, abrupt changes in $|S_\mathrm{11}|$ appear, most prominently in Fig.~\ref{fig:VNA}~(f). These discontinuities indicate that, at specific combinations of drive power and magnetic field, the microwave response of the coupled system changes abruptly. Related  phenomena have previously been reported in strongly driven magnon-polariton systems based on the $k=0$ ferromagnetic resonance mode coupled to an electromagnetic cavity \cite{PhysRevLett.120.057202}. In the present case, the microscopic origin of these discontinuities cannot be determined from the VNA data alone and is therefore addressed in Secs.~\ref{sec:BLS} and \ref{sec:SA} using complementary measurements.

Overall, the VNA measurements establish the key phenomenology of the driven SAW-SW system: strong coupling in the linear regime, power-induced broadening and reduced contrast of the hybridized modes, a systematic shift of the coupling features, and the emergence of discontinuities in the electrical response. The physical origin of these effects is clarified in the following sections.

\subsection{Nonlinear frequency shift}\label{sec:Nonlinear}

The VNA measurements show that, with increasing drive power, the resonance shifts to lower magnetic fields. This behavior implies a positive nonlinear frequency shift of the SAW-driven spin-wave mode. At first sight, this is unexpected, since in in-plane magnetized YIG films the nonlinear frequency self-shift is often found to be negative for a broad range of thicknesses and wave vectors \cite{Bunkov2023InverseFoldoverYIG, Lake2022NonlinearSpectralShiftYIG, Schneider2020-xz}. In YIG spheres, the nonlinear frequency shift can be dominated by the magnetocrystalline anisotropy because the spherical geometry does not give rise to a shape anisotropy \cite{PhysRevLett.120.057202}. In the present film geometry, however, this contribution is expected to be secondary. The characteristic energy density associated with the cubic anisotropy of YIG is only $|K_1|\approx \qty{6e2}{\joule\per\meter\cubed}$ $(|K_2|\ll|K_1|)$\cite{Hansen1974}, whereas the demagnetization-induced shape anisotropy of our film is approximately of the order $\mu_0 M_s^2/2\approx 1\times10^4\,\mathrm{J\,m^{-3}}$ \cite{Osborn1945}.

In the present system, the nonlinear response is not determined by the self-shift alone. The standing SAW cavity mode contains counterpropagating components and therefore couples simultaneously to the $+k$ and $-k$ SW states. Consequently, the frequency of either driven mode is affected both by a nonlinear self-shift due to its own occupation and by a nonlinear cross-shift induced by the occupation of the counterpropagating mode.

To quantify these contributions, we employ the vector Hamiltonian formalism for nonlinear spin-wave dynamics \cite{WagnerInPreparation, tyberkevych2020vector}, which extends the standard Hamiltonian formalism for spin waves \cite{krivosik_hamiltonian_2010}. This framework has previously been applied successfully to nonlinear phenomena in spin-wave waveguides \cite{Wang2023-qv}. Within this approach, the nonlinear frequency shift is described by effective interaction coefficients. The complex amplitudes of the two counterpropagating SWs are denoted by $c_\mathrm{\pm k}$, with $\lvert c_\mathrm{\pm k}(t)\rvert^2$ proportional to the corresponding classical magnon occupation numbers. In the symmetric-cavity approximation, the reciprocal SW dispersion and equal magnitudes of the magnetoelastic coupling to the two counterpropagating components motivate the assumption $\lvert c_{+k}\rvert^2=\lvert c_{-k}\rvert^2\equiv\lvert c_\mathrm{S}\rvert^2$ \cite{natcom2025}. The shifted frequency $\widetilde{f}_\mathrm{k}$ of the $k$ mode can then be written as

\begin{equation}\label{eq:Nonlinshift}
    \begin{split}
        \widetilde{f}_\mathrm{k}
        &=
        f_\mathrm{k}
        +
        W_\mathrm{kk,kk}\lvert c_\mathrm{k}\rvert^2
        +
        2W_\mathrm{-kk,-kk}\lvert c_\mathrm{-k}\rvert^2\\
        &=
        f_k+
        \left(W_\mathrm{kk,kk}+2W_\mathrm{-kk,-kk}\right)
        \lvert c_\mathrm{S}\rvert^2 .
    \end{split}
\end{equation}

Here, $W_\mathrm{kk,kk}$ describes the effective nonlinear self-shift of the $k$ mode, whereas $2W_{-kk,-kk}$ describes the effective cross-shift induced by the counterpropagating mode.

\begin{figure*}[tbp]
\centering
\includegraphics[width=\textwidth]{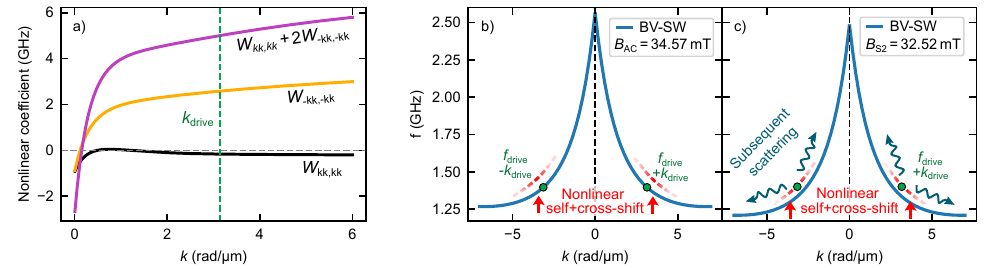}
\caption{(a) Nonlinear self-shift coefficient $W_\mathrm{kk,kk}$, cross-shift coefficient $W_\mathrm{-kk,-kk}$, and total nonlinear-shift coefficient $W_\mathrm{tot}=W_\mathrm{kk,kk}+2W_\mathrm{-kk,-kk}$ as functions of the wave vector $k$. The green dashed line marks the wave vector $k_\mathrm{drive}$ of the driving phonon, which is resonantly coupled to the magnon in the linear regime. (b,c) Calculated linear dispersion of the BV-SW mode (blue). The green circles indicate the SAW-driven states. In (b), the drive is resonant with the BV-SW in the linear regime. With increasing power, the positive nonlinear shift detunes the SW from the fixed drive frequency, as indicated by the red arrows and dashed red lines. In (c), the drive is initially off-resonant (lower $B_\mathrm{ext}$) and is shifted into resonance by the positive total nonlinear frequency shift. Once the resonance condition is met at larger driving powers, the excitation can become instable, after which additional scattering channels open (schematically indicated by dark-blue arrows).}
\label{fig:Nonlinear}
\end{figure*}

Figure~\ref{fig:Nonlinear}~(a) shows the calculated self-shift and cross-shift coefficients together with the resulting total coefficient $W_\mathrm{tot}=W_\mathrm{kk,kk}+2W_\mathrm{-kk,-kk}$ at $B_\mathrm{AC}=\qty{34.57}{\milli\tesla}$. The remaining parameters used in the calculation are listed in the Sec.~S3.1 of the supplemental material \cite{SupplementalMaterial}. The green dashed line marks the wave vector of the magnon mode driven by the phonon. Under the assumption of equal occupation of the counterpropagating modes, the resulting total nonlinear-shift coefficient at $k_\mathrm{drive}$ is $W_\mathrm{tot}=\qty{5.00}{\giga\hertz}$.

The key elements of the theoretical model relevant to the present calculations are summarized in Sec.~S3 of the Supplemental Material \cite{SupplementalMaterial}, which includes Refs.~\cite{krivosik_hamiltonian_2010, tyberkevych2020vector, MagOscillations_Gurevich1996, GreensFunction_Guslienko2011}, while a detailed derivation is provided in Ref.~\cite{WagnerInPreparation}.

Interestingly, the predicted positive frequency shift does not require equal populations of the counterpropagating modes. At $k_\mathrm{drive}$, the effective cross-shift exceeds the self-shift in magnitude, $2|W_{-kk,-kk}|>|W_{kk,kk}|$, such that the positive cross-nonlinearity dominates over a broad range of population asymmetries. The net nonlinear frequency shift therefore remains positive even for significantly unequal populations. At a fixed drive frequency, this positive shift displaces the resonance toward lower magnetic fields, consistent with the VNA measurements shown in Fig.~\ref{fig:VNA}.

In Fig.~\ref{fig:Nonlinear}~(b) and \ref{fig:Nonlinear}~(c), the dispersion relations of the BV-SW are shown for $B_\mathrm{AC}$ and $B_\mathrm{S2}$. In panel b), the SAW drive at $f_\mathrm{drive}$ and $\pm k_\mathrm{drive}$, indicated by the green points, is resonant with the BV-SW mode at low driving power. As the driving power is increased, the nonlinear self- and cross-shift progressively detune the spin-wave dispersion from the fixed SAW drive, as schematically illustrated by the red dashed line. As a consequence, the SAW-SW coupling efficiency is expected to decrease. This behavior is discussed in more detail in Fig.~\ref{fig:SA}~(a).

In contrast, when the magnetic field in the linear regime is set below the linear anticrossing field $B_\mathrm{AC}$, the system is expected to exhibit a qualitatively different behavior. This case is illustrated in Fig.~\ref{fig:Nonlinear}~(c), where $B_\mathrm{S2}$ is approximately $\qty{2}{\milli\tesla}$ smaller than $B_\mathrm{AC}$. Here, the fixed SAW drive is initially detuned from the BV-SW dispersion. However, due to the positive total nonlinear shift, the BV-SW dispersion at $\pm k_\mathrm{drive}$ is shifted towards the drive frequency with increasing power, such that resonance is reached at higher drive powers.

This leads to foldover behavior of the resonance curve, with an abrupt increase in SW intensity once the driving power is sufficient to sustain the amplitude required to shift the mode into resonance \cite{Jiang2025-jl, Wang2024-su, PhysRevLett.120.057202, Guo2026-eg, Geilen2025-ai, Breitbach2026-wq, Guo2026-br}. To assess whether the calculated magnon nonlinearity can account for the observed foldover, we employ a reduced driven coupled-mode model incorporating SAW-SW coupling, damping, detuning, and the nonlinear magnon frequency shift. Although the phonon mode provides the dominant excitation channel, it is treated as linear in this model, such that the nonlinearity originates entirely from the spin-wave subsystem. The model qualitatively reproduces the field-dependent foldover and hysteresis observed experimentally; its formulation and calculated responses are presented in Sec.~S7 of the supplemental material.  In supplement Sec.~S5, which includes Refs. \cite{Hwang2023, Spindler2013, Wang2015ZnO}, we additionally examine alternative sources of frequency shifts, including heating and an intrinsic power-dependent shift of the uncoupled SAW mode, and find that their contributions are too small to account for the experimental observations. Together, these results support magnon-magnon interactions as the leading origin of the observed nonlinear response within the scope of the present analysis.

At an applied microwave power of $P_\mathrm{MW}=\qty{10}{\dBm}$, the foldover occurs at approximately $\qty{31.5}{\milli\tesla}$ [Fig.~\ref{fig:SA}~(g)], corresponding to a displacement of $\qty{3.07}{\milli\tesla}$ towards lower fields relative to the linear-resonance field $B_\mathrm{AC}=\qty{34.57}{\milli\tesla}$. Using the calculated field-dependent SW dispersion at $k_\mathrm{drive}$, this displacement corresponds to an estimated nonlinear frequency shift of $\qty{89}{\mega\hertz}$. With the total nonlinear shift coefficient of $\qty{5.00}{\giga\hertz}$ and assuming equal populations of the counterpropagating modes, Eq.~\eqref{eq:Nonlinshift} yields a SW amplitude of $|c_\mathrm{S}|\approx0.13$. This value is comparable in magnitude to amplitudes reported for nonlinear SWs driven directly by microwave antennas \cite{Wang2024-su, Breitbach2026-wq}.

If the SW intensity on this nonlinear resonance branch exceeds the threshold for instabilities induced by magnon-magnon scattering \cite{Kreil2018-iy}, additional scattering channels open, as schematically indicated by the dark-blue arrows in Fig.~\ref{fig:Nonlinear}~(c). These processes redistribute energy across the magnon spectrum. The resulting behavior is investigated experimentally in the following sections.

\subsection{Nonlinear dynamics observed by $\upmu$BLS}\label{sec:BLS}

To gain insight into the dynamics at the quasiparticle level, we employ $\upmu$BLS spectroscopy. Unlike the electrical measurements, this optical technique does not rely on the IDT for SAW-to-electrical-signal conversion and is therefore not restricted to the working-frequency of the IDT. Instead, $\upmu$BLS spectroscopy allows direct detection of the quasiparticle frequency $f_\mathrm{BLS}$ via the photon frequency shift induced by the scattering process. Moreover, because phonon and magnon scattering are associated with different polarization rotations \cite{Sebastian2015-hs, Geilen2022-zo}, the two contributions can be distinguished. In our system, the phonon signal is suppressed by approximately 96.5\% in the magnon polarization channel (detailed in Sec.~S1 of the supplement), and the remaining residual phonon contribution is subtracted from the measured magnon signal.

\begin{figure}[tbp]
    \centering
    \includegraphics[width=\columnwidth]{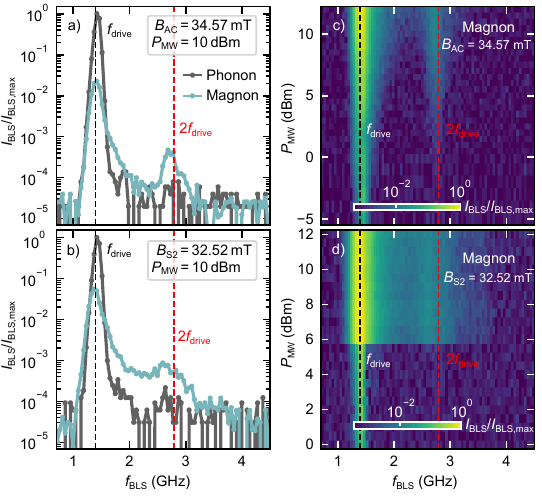}
    \caption{Observation of nonlinear processes using $\upmu$BLS spectroscopy. (a,b) Normalized $\upmu$BLS intensity $I_\mathrm{BLS}/I_\mathrm{BLS,max}$ as a function of $f_\mathrm{BLS}$ measured in magnon and phonon polarization at $P_\mathrm{MW}=\qty{10}{\dBm}$ for two magnetic fields. The black and red dashed lines mark $f_\mathrm{drive}$ and $2f_\mathrm{drive}$, respectively. In (a), the field is tuned to the central anticrossing in the linear regime, and a nonlinear magnon signal is observed at $2f_\mathrm{drive}$. In (b), the field is below the linear anticrossing field, resulting in a broad magnon spectrum, while a phonon signal is only present at $f_\mathrm{drive}$ in both cases. (c,d) $I_\mathrm{BLS}/I_\mathrm{BLS,max}$ as a function of $P_\mathrm{MW}$ and $f_\mathrm{BLS}$. In (c), resonant excitation at the anticrossing field leads to the onset of scattering to $2f_\mathrm{drive}$ at powers above \qty{0}{\dBm}. In (d), the low-power response at $f_\mathrm{drive}$ is initially forced and becomes resonant due to the nonlinear frequency shift at approximately $\qty{6}{\dBm}$, above which a broad magnon spectrum emerges.}
    \label{fig:BLS}
\end{figure}

Figure~\ref{fig:BLS} shows the $\upmu$BLS measurements acquired at $B_\mathrm{AC}$ and $B_\mathrm{S2}$. The $\upmu$BLS intensity is normalized independently in each panel to the largest value displayed: single-signal spectra to their respective maximum, combined spectra to the common maximum of both signals, and two-dimensional maps to their global maximum. Consequently, relative intensities can be compared within, but not between, individual panels.

We first consider the upper row, measured at \mbox{$B_\mathrm{AC}=\qty{34.57}{\milli\tesla}$}. Panel~\ref{fig:BLS}~(a) displays the normalized $\upmu$BLS intensity $I_\mathrm{BLS}/I_\mathrm{BLS,max}$ as a function of the quasiparticle frequency $f_\mathrm{BLS}$ for phonon and magnon signals at a high microwave power of $P_\mathrm{MW}=\qty{10}{\dBm}$. The drive frequency is chosen to match the central high-$Q$ SAW mode of the resonator, $f_\mathrm{drive}=\qty{1397.6}{\mega\hertz}$; this frequency is also indicated in the VNA measurements shown above in Fig.~\ref{fig:VNA}~(c,e,g) by the red dashed line. 

At this field, the phonon signal is observed only at $f_\mathrm{drive}$, whereas magnons are detected at both $f_\mathrm{drive}$ and $2f_\mathrm{drive}$. Figure~\ref{fig:BLS}~(c) shows the normalized magnon intensity $I_\mathrm{BLS}/I_\mathrm{BLS,max}$ as a function of $f_\mathrm{BLS}$ and $P_\mathrm{MW}$. The magnon signal near $2f_\mathrm{drive}$ emerges gradually with increasing power, with an onset already at approximately $\qty{0}{\dBm}$.

The calculated dispersion in Fig.~\ref{fig:Nonlinear}~(b) shows that the lowest-order BV-SW branch provides no state at $2f_\mathrm{drive}$ that simultaneously satisfies energy and in-plane momentum conservation to enable a resonant scattering process. The signal at $2f_\mathrm{drive}$ may therefore originate either from a forced, nonresonant magnetization response \cite{Lvov2023-qo,Geilen2025-ai} or from a weak three-magnon confluence process involving higher-order PSSW modes. In the latter case, the rapidly oscillating thickness profiles of the PSSW modes substantially reduce their overlap with the fundamental SAW-driven mode, such that the confluence process is expected to be inefficient. This is discussed in detail in Sec.~S4 of the supplemental material, which includes Ref.~\cite{Etesamirad2023}.

As the magnon population at $f_\mathrm{drive}$ continues to increase, the nonlinear self- and cross-shifts progressively detune the excitation, as illustrated schematically in Fig.~\ref{fig:Nonlinear}~(b) by the red arrow and dashed line. Consequently, the coupling between the SAW and the directly driven magnon mode is expected to decrease. This is indeed observed in Fig.~\ref{fig:SA}~(a), and discussed in Sec.~\ref{sec:SA}.

The lower row of Fig.~\ref{fig:BLS} shows the situation at a magnetic field below the anticrossing field of the linear regime $B_\mathrm{S2}$. In Fig.~\ref{fig:BLS}~(b), obtained at $P_\mathrm{MW}=\qty{10}{\dBm}$, the magnon signal is distributed over a broad frequency range, whereas the phonon signal remains at $f_\mathrm{drive}$ as before. Although the normalized logarithmic representation may suggest a weak phonon feature near $2f_\mathrm{drive}$, repeated measurements show that it lies at the noise level and is not reproducible.The color-coded map of $I_\mathrm{BLS}/I_\mathrm{BLS,max}$ in Fig.~\ref{fig:BLS}~(d) further shows that at a critical power of about $\qty{6}{\dBm}$, the magnon intensity exhibits a jump and extends over a broad frequency range, with a local maximum near $2f_\mathrm{drive}$.

To understand this different behavior, we come back to the calculated dispersion in Fig.~\ref{fig:Nonlinear}~(c) for this lower external field. In this case, the purple dots marking the SAW-driven states at $f_\mathrm{drive}$ and $\pm k_\mathrm{drive}$ do not satisfy the resonance condition in the low-power regime. Nevertheless, a forced excitation at $f_\mathrm{drive}$ is still present, as visible in Fig.~\ref{fig:BLS}~(d) for powers below $\qty{6}{\dBm}$. Due to this forced excitation, the magnon population increases with growing $P_\mathrm{MW}$ until the positive nonlinear frequency shift becomes large enough to bring the magnon into resonance with the phonon. This is schematically indicated in Fig.~\ref{fig:Nonlinear}~(c) by the small red arrows and dashed lines near the purple excitation points, representing the dispersion under consideration of nonlinear interactions.

Once this critical power is reached, the population of the driven magnons at $\pm k$ and $f_\mathrm{drive}$ increases abruptly, and, as a secondary effect, the resonant response at $f_\mathrm{drive}$ becomes dynamically instable. As a result, a cascade of secondary scattering processes, including four-magnon processes involving thermally populated magnons \cite{Geilen2025-ai}, is triggered and populates a broad part of the accessible spin-wave band. This is illustrated schematically in Fig.~\ref{fig:Nonlinear}~(c) by the dark-blue arrows and is supported by the measurement in Fig.~\ref{fig:BLS}~(d). The lowest frequency visible in panel~(d) is in good agreement with the calculated band-bottom frequency at roughly $k=\qty{\pm 6}{\radian\per\micro\meter}$. Forced excitations or a weak confluence process at $2f_\mathrm{drive}$ still occur, but they are not the primary processes contributing to the observed spectrum. Within the framework of the existing literature, such a broad redistribution of spectral weight is consistent with a spin-wave instability and may also be related to bistable foldover \cite{PhysRevLett.120.057202, Wang2024-su, Chai2019-ci, Kreil2018-iy, Guo2026-br, Jiang2025-jl}. A more conclusive discussion of the latter requires hysteretic power sweeps which are presented in \ref{sec:SA}.

For completeness, we note that for magnetic fields above $B_\mathrm{AC}$, the corresponding $\upmu$BLS spectra shown in Sec.~S2 of the supplement exhibit only a weak forced excitation. In this regime, the spin-wave dispersion is shifted to higher frequencies, i.e., further away from resonance with the SAW drive, and the positive nonlinear frequency shift therefore does not bring the system into resonance.

Overall, the $\upmu$BLS data show that the nonlinear response is governed by the initial resonance detuning set by the magnetic field. When the system is resonant in the linear regime, the nonlinear frequency shift progressively detunes the coupling with increasing power, whereas under initially off-resonant conditions an instability emerges once resonance is reached at higher power. This behavior is also predicted by the coupled magnon-phonon model presented in Sec.~S7 of the supplement.

\subsection{Bistability}\label{sec:SA}

Having established that the threshold-like response of the system originates from the nonlinear magnon frequency shift and the associated overpopulation of the magnon mode at $f_\mathrm{drive}$, we now demonstrate that the system shows a bistable foldover. To this end, we compare hysteretic microwave-power sweeps measured by $\upmu$BLS with the corresponding electrical response recorded by the spectrum analyzer (SA). The main goal of this section is therefore twofold: first, to verify that the observed jumps are hysteretic, as expected for a bistable foldover \cite{Bunkov2023InverseFoldoverYIG, PhysRevLett.120.057202, Gui2009-we, Wang2024-su}, and second, to use the electrical readout as a fast probe to map the bistability over a wide range of magnetic field and drive power.

As in the previous section, the SAW resonator is driven at a fixed frequency of $f_\mathrm{drive}=\qty{1397.6}{\mega\hertz}$. From the SA measurement we extract the detected power $P_\mathrm{SA}$ and define
\begin{equation}\label{eq:alpha}
    \alpha_\mathrm{up/down}=\frac{\mathrm{d}(P_\mathrm{SA} (\mathrm{dBm)})}{\mathrm{d}(P_\mathrm{MW,up/down}(\mathrm{dBm)})}~,
\end{equation}
where the subscripts up/down denote increasing and decreasing microwave power sweeps, respectively. The quantity $\alpha$ highlights abrupt changes in the electrical response of the IDT-cavity system and therefore provides a sensitive measure of small deviations of the phonon signal.

This electrical data is compared to the normalized integrated $\upmu$BLS intensity, $\int I_\mathrm{BLS}/\int I_\mathrm{BLS,max}$. Specifically, we evaluate three contributions: the phonon intensity at $f_\mathrm{drive}$, the magnon intensity at $f_\mathrm{drive}$, and the magnon intensity at $2f_\mathrm{drive}$. For each contribution, the $\upmu$BLS signal is integrated over a frequency range covering the respective spectral feature. The ranges are presented in Sec.~S6 of the supplement. As shown in Fig.~\ref{fig:BLS}~(a,b), no phonon signal is observed at $2f_\mathrm{drive}$; consequently, no corresponding trace is displayed. Within each panel, all integrated traces are normalized by a common factor $\int I_\mathrm{BLS,max}$, defined as the largest integrated intensity among the traces displayed in that panel.

\begin{figure*}[tbp]
    \centering
    \includegraphics[width=\textwidth]{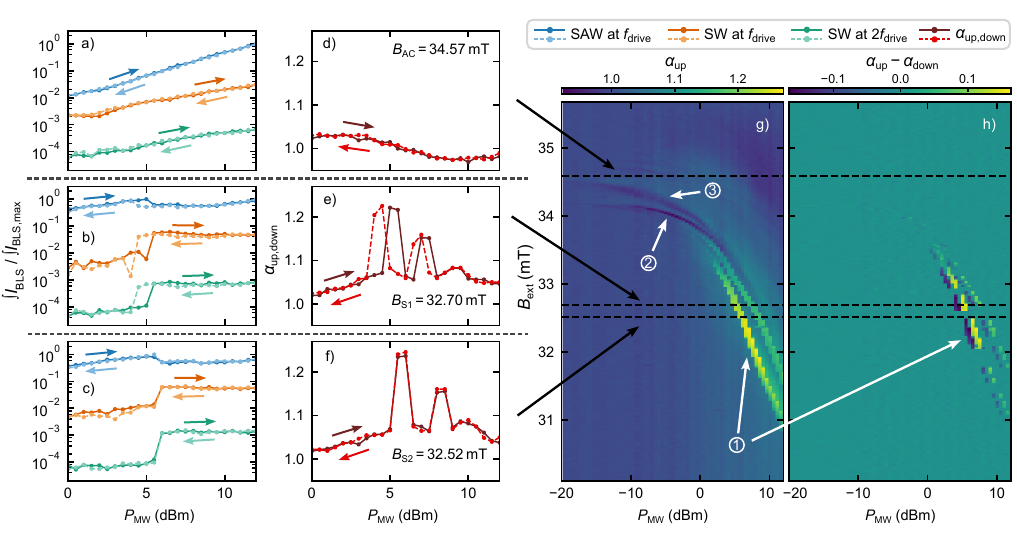}
    \caption{
        Bistability in the coupled magnon-phonon system revealed by hysteretic $P_\mathrm{MW}$ sweeps in $\upmu$BLS and SA measurements. 
        (a-c) Normalized integrated $\upmu$BLS intensities, $\int I_\mathrm{BLS}/\int I_\mathrm{BLS,max}$, for the phonon mode at $f_\mathrm{drive}$ and the magnon modes at $f_\mathrm{drive}$ and $2f_\mathrm{drive}$, measured during upward and downward $P_\mathrm{MW}$ sweeps at the indicated magnetic fields. 
        (d-f) Corresponding spectrum-analyzer response, expressed as $\alpha_\mathrm{up/down}=\mathrm{d}P_\mathrm{SA}/\mathrm{d}P_\mathrm{MW,up/down}$. 
        At the anticrossing field, $B_\mathrm{AC}=\qty{34.57}{\milli\tesla}$, all signals evolve continuously and no pronounced hysteresis is observed [panels (a) and (d)]. 
        At $B_\mathrm{S1}=\qty{32.70}{\milli\tesla}$, a clear bistability appears: the phonon intensity at $f_\mathrm{drive}$ drops abruptly, while the magnon intensities at $f_\mathrm{drive}$ and $2f_\mathrm{drive}$ increase, accompanied by hysteretic spikes in the electrical response [panels (b) and (e)]. 
        At $B_\mathrm{ext}=\qty{32.52}{\milli\tesla}$, a jump is still present, but no pronounced hysteresis is resolved [panels (c) and (f)]. 
        (g) Field- and power-dependent map of $\alpha_\mathrm{up}$. The dashed black horizontal lines indicate the magnetic fields shown in panels (a-f). The features labeled 1-3 denote distinct nonlinear features of different origin, as discussed in the text. 
        (h) Corresponding map of $\alpha_\mathrm{up}-\alpha_\mathrm{down}$, highlighting the regions in which hysteresis is observed. The most pronounced bistability is associated with feature 1. The interruptions along this hysteretic branch are discussed in the main text.
        }
    \label{fig:SA}
\end{figure*}

Figure~\ref{fig:SA} is organized as follows: panels (a-f) show representative up- and down-sweeps at three selected magnetic fields, measured by $\upmu$BLS (a-c) and SA (d-f), while panels (g) and (h) display the full field- and power-dependent maps of $\alpha_\mathrm{up}$ and $\alpha_\mathrm{up}-\alpha_\mathrm{down}$, respectively. The horizontally adjacent panels correspond to the same magnetic field. 

We first consider the field $B_\mathrm{AC}=\qty{34.57}{\milli\tesla}$, which corresponds to the anticrossing field in the low-power, linear regime. The integrated $\upmu$BLS data shown in Fig.~\ref{fig:SA}~(a) exhibit a continuous increase of all three signals with increasing power. However, the phonon signal at $f_\mathrm{drive}$ increases more strongly than the two magnon contributions. This behavior is consistent with the positive nonlinear magnon frequency shift: as the spin-wave dispersion shifts upward in frequency, the driven state becomes increasingly detuned from the spin-wave branch, which reduces the efficiency of phonon-to-magnon conversion. Importantly, no qualitative difference between the up- and down-sweeps is observed. The corresponding SA data in Fig.~\ref{fig:SA}~(d) likewise shows no pronounced change in $\alpha$, indicating the absence of a foldover at this field.

A qualitatively different behavior is found slightly below the anticrossing field, at $B_\mathrm{S1}=\qty{32.70}{\milli\tesla}$. The $\upmu$BLS data in Fig.~\ref{fig:SA}~(b) displays a clear threshold like feature in all three integrated signals. At a critical microwave power, the phonon intensity at $f_\mathrm{drive}$ drops abruptly, while the magnon intensities at both $f_\mathrm{drive}$ and $2f_\mathrm{drive}$ increase. This behavior is consistent with the mechanism established in the previous section: due to the positive nonlinear frequency shift, the spin-wave mode is shifted into resonance with the drive as the power is increased. Once this condition is reached, energy is transferred resonantly from the phonon to the magnon system, leading to a rapid overpopulation of the driven magnon mode and the opening of additional scattering channels. Overall, this results in a reduced phonon population and an enhanced magnon population. The simultaneous increase of the magnon intensities at $f_\mathrm{drive}$ and $2f_\mathrm{drive}$ is further consistent with Fig.~\ref{fig:BLS}~(d), which shows that, at the same critical power, the magnon population broadens over a wide frequency range due to scattering processes originating from the excitation at $f_\mathrm{drive}$. Most importantly, the transition occurs at different powers for upward and downward power sweeps, demonstrating a pronounced hysteresis. This behavior strongly supports the interpretation in terms of a foldover bistability of the driven spin-wave mode at $f_\mathrm{drive}$.

A notable feature emerges beyond the foldover transition, where all three spectrally selected signals exhibit a reduced slope with increasing $P_\mathrm{MW}$. To determine whether this behavior merely reflects a redistribution of spectral weight, we additionally integrate the magnon and phonon intensities over their full measured frequency ranges, as detailed in Sec.~S6 of the supplement. The resulting total magnon and phonon spectral weights show the same reduced power dependence, indicating that the coupled response stabilizes beyond the foldover transition.

The SA data shown in Fig.~\ref{fig:SA}~(e) is in agreement with the $\upmu$BLS results. The quantity $\alpha$ exhibits pronounced spikes at different powers for the up- and down-sweeps, directly reflecting the bistable transition seen optically. However, in contrast to the $\upmu$BLS data, the SA trace resolves two spikes rather than one, analogous to the two-step structure already observed in the VNA response in Fig.~\ref{fig:VNA}~(f). We return to this point below.

A third representative field, $B_\mathrm{S2}=\qty{32.52}{\milli\tesla}$, is shown in Fig.~\ref{fig:SA}~(c,f). Here, both $\upmu$BLS and SA again exhibit a clear jump, confirming that the nonlinear instability remains present. Unlike the case discussed above, no pronounced separation between the up- and down-sweeps is resolved in either the optical or the electrical data. This demonstrates that the occurrence of the jump itself does not necessarily imply a strongly developed bistable window, as will be discussed below.

Having established that the SA response tracks the identical instabilities/bistabilities identified by $\upmu$BLS, we can now use the electrical readout to map these features over a wide field and power range, which would be prohibitively time consuming by $\upmu$BLS alone. The resulting map of $\alpha_\mathrm{up}$ is shown in Fig.~\ref{fig:SA}~(g), where the dashed horizontal lines mark the three field values discussed above. Three further distinct features, labeled 1-3 and marked with white arrows, can be identified (curved lines shifting to smaller magnetic fields with increasing power).

The most important signal is feature 1, which corresponds to the foldover discussed above and in Sec.~\ref{sec:BLS}. This branch marks the jump associated with the overpopulation of the magnon mode at $f_\mathrm{drive}$ and therefore provides a direct map of the nonlinear transition in the $B_\mathrm{ext},P_\mathrm{MW}$ plane. In this way, we can efficiently track the field and power dependence of the foldover using a SA.

Feature 2 is less pronounced and remains visible over a broad power range. Its origin can be traced back to the olive-marked region of the VNA data in Fig.~\ref{fig:VNA}~(b), where it corresponds to the dip at slightly lower magnetic field associated with the upper polaron branch of the lower-frequency high-$Q$ SAW mode crossing $f_\mathrm{drive}$ \cite{natcom2025}. This branch is expected to exhibit behavior similar to feature 1, as it is driven by the same mechanism and differs mainly in arising from the hybridized branch of a slightly shifted mode. Its influence on the main physics discussed here is, however, negligible: the $\upmu$BLS data in Fig.~\ref{fig:SA}~(a-c) show no significant change at the corresponding power. We therefore do not consider this branch further.

Feature 3 is the weakest signal and is attributed to the directly driven spin-wave excitation induced by spurious currents in the microstructures \cite{natcom2025}. At low powers, it appears at the same magnetic field as the second dip in the olive-marked region of the VNA data in Fig.~\ref{fig:VNA}~(b), consistent with its identification as the directly driven spin-wave resonance. With increasing power, this feature shifts due to the nonlinear frequency shift, but remains weak and becomes difficult to resolve above approximately $\qty{0}{\dBm}$.

The hysteresis map shown in Fig.~\ref{fig:SA}~(h), where the color scale represents $\alpha_\mathrm{up}-\alpha_\mathrm{down}$, provides a direct measure of the bistable regions. The strongest hysteresis is associated with feature 1, consistent with its interpretation as a foldover bistability of the resonantly driven magnon mode. At the same time, the map shows that the bistability is not equally pronounced along the entire branch. For example, at $B_\mathrm{S2}=\qty{32.52}{\milli\tesla}$, a clear jump is still observed, whereas no pronounced hysteresis can be resolved. A possible explanation is that the change in magnetic field modifies the efficiency of the secondary scattering channels and may also change which modes participate in the scattering process. This would alter the effective nonlinear relaxation and damping experienced by the driven mode \cite{Scott2004-tj}, thereby affecting the width and visibility of the bistability region. Furthermore, if the bistable window becomes narrower than the experimental power resolution of \qty{0.5}{\dBm}, the bistability would remain unresolved in the present experiment.

Overall, the comparison between $\upmu$BLS and SA measurements demonstrates that the dominant electrically detected peak is a direct signature of the nonlinear magnon-phonon dynamics resolved optically. In particular, the onset of bistability has a pronounced impact on the electrically detected phonon response. A notable finding is that, beyond the foldover transition, not only the magnon signals but also the phonon response exhibits a drastically reduced slope as a function of $P_\mathrm{MW}$, indicating that the phonon mode also stabilizes. More generally, the observed stabilization of the phonon response mediated by the nonlinear magnon dynamics may open new perspectives for controlling phonon amplitudes in magnetoelastic devices. In Sec.~S7 of the supplement, we further demonstrate that a reduced coupled magnon-phonon model reproduces the foldover bistability observed in our system.

\section{Conclusion}

In conclusion, we have investigated the nonlinear dynamics of a strongly coupled magnon-polaron system. Our electrical and spectroscopic measurements, together with calculations based on the vector Hamiltonian formalism, support a positive nonlinear frequency shift as the principal mechanism underlying the observed field- and power-dependent response. Within this model, the total shift is dominated by the cross-shift arising from the simultaneous excitation of counterpropagating SWs at $+k$ and $-k$ by the standing SAW cavity mode. Although our calculation includes only the nonlinear dynamics of the spin-wave subsystem it reproduces the sign and qualitative field dependence of the observed shift. 

Furthermore, we have mapped the field and power ranges over which foldover and bistability occur and directly related the macroscopic electrical response to the underlying magnon dynamics. The correspondence between the electrical and $\upmu$BLS measurements indicates that the foldover is accompanied by a redistribution of the magnon population across a broad spin-wave spectrum. Beyond the foldover threshold, both the spin-wave and SAW intensities no longer increase proportionally with microwave power, indicating a limiting behavior governed by nonlinear scattering processes.

Our results also point to further opportunities offered by SAW-driven magnonic systems. The field- and frequency-dependent onset of the instability provides controlled access to a strongly nonlinear regime in which a broad magnon spectrum is populated. Such excitation may offer a route toward high-magnon-density regimes relevant to collective phenomena such as magnon Bose-Einstein condensation. More broadly, the investigated system provides a promising platform for studying nonlinear magnon-phonon interactions and engineered spin-wave scattering processes.

\begin{acknowledgments}
This work was supported by the European Research Council (ERC) under the European Union’s Horizon Europe research and innovation programme (Consolidator Grant ”MAWiCS”, Grant Agreement No. 101044526), the Deutsche Forschungsgemeinschaft (DFG, German Research Foundation) within the Transregional Collaborative Research Center TRR 173/3-268565370 Spin+X (Project B01) and by the ERC Starting Grant "CoSpiN" 10104243.
KL acknowledges support from a Clarendon Fund Scholarship.
\end{acknowledgments}

\section*{Author contributions}
M.We. conceived the project. K.K., K.L., and J.-F.G. fabricated the sample. K.K. carried out the measurements. P.K., M.-R.S., and E.S. assisted with the experimental work. M.Wa. performed the theoretical calculations. K.K. analyzed and interpreted the results in discussion with M.Wa., Y.K., P.P., and M.We. K.K. wrote the manuscript with input from all coauthors.

\section*{Competing interests}
The authors declare that they have no competing interests.

\section*{Data availability}
The data supporting the findings of this study is available on Zenodo \cite{ZenodoData}.

\bibliography{references.bib} 

\end{document}